# Phenomenological BCS theory of the high-$T_c$ cuprates

R. Fehrenbacher and M.R. Norman

*Materials Science Division,*
*Argonne National Laboratory,*
*Argonne, Illinois 60439*

**Abstract**. A BCS model characterized by a phenomenological pair potential with on-site ($V_0$), nearest ($V_1$), and next nearest ($V_2$) neighbour coupling constants, and an empirical quasiparticle dispersion taken from angle-resolved photoemission spectra is considered. The model can consistently explain the experimental data concerning the pair state of the hole doped cuprates. Three ingredients are required to make the interpretation possible: the existence of flat bands, a very small effective on-site repulsion, and a slightly dominating effective nnn attraction $V_2$ of the order of 60-80meV with a ratio $V_2/V_1 \approx 1.5$.

A number of recent experiments[1, 2] have indicated that the order parameter (OP) of the hole doped high-$T_c$cuprates has a strong **k** dependence, and is very likely exhibiting line nodes. Moreover, a subset of these experiments[3], all of them using $YBa_2Cu_3O_{7-y}$(Y123) samples, provided strong direct evidence that the OP changes sign under a 90° rotation, reminiscent of $d_{x^2-y^2}$-symmetry. However, another set of experiments on Y123[4] which also directly probe the **k** dependence of the OP show results that are hard to reconcile assuming a pure $d_{x^2-y^2}$ gap. Also, the temperature dependence of the London penetration depth measured on crystals of the electron doped $Nd_{1.85}Ce_{0.15}CuO_{4-y}$(NCCO) is consistent with a weakly anisotropic BCS-like energy gap[5].

More recently, new information on the **k** dependence of the superconducting gap of high quality $Bi_2Sr_2CaCu_2O_8$ (Bi2212) crystals was reported from angle-resolved photoemission (ARPES) data by the Argonne group[6]. The gap on the Fermi surface (FS) was found to vanish at *two* points symmetrically displaced *away* from the (1,1) direction, but was finite on the (1,1) line. This result is incompatible with a $d_{x^2-y^2}$-OP, which must have a *single* node along the (1,1) direction (a small admixture of an s-wave component, could slightly displace the node to one side, but cannot produce two nodes)[7]. Instead, the angular dependence of the gap resembled very closely that of an OP of the form $\Delta_{\mathbf{k}} \sim \cos k_x \cos k_y$ [8], which we shall call the $s_{xy}$ state.

Considering these findings, one might draw the unpleasant conclusion that the pairing mechanism of the cuprates might be non-universal. A possible alternative was proposed by Chakravarty *et al.* [9]. In their theory, the main ingredient to the high $T_c$ is the postulate that the normal state of the cuprates is a Luttinger liquid confined to single $CuO_2$-sheets. This implies *non-coherent single-particle* charge transport in the *c*-direction. In the superconducting state, however, *coherent pair* tunneling becomes possible, and the corresponding large gain in kinetic energy leads to the high $T_c$. Within this scenario, the symmetry of the pair state is determined by weak residual interactions within a single layer, and can thus easily vary between different materials.

In this article, we make an attempt to find a consistent interpretation of the data within a simple BCS framework. We construct a phenomenological model that can qualitatively account for the apparently different pair states. The experimental quasiparticle dispersion relation, as determined by ARPES, serves as an input to the theory, and for the pair potential, we allow on-site, nearest (nn), and next nearest neighbour (nnn) interactions.

The main finding of the theory is the following: the degree to which the Cooper pairs can take advantage of the large single particle density of states (DOS) (arising from the observed flat bands in the quasiparticle dispersion of hole doped cuprates [10]), highly depends on the **k** dependence of the pair wavefunction. Assuming singlet pairing, and excluding the isotropic *s*-wave case by using a repulsive on-site interaction, we shall show that due to this effect, for the hole doped materials, there are precisely *two* different competing pair states, namely the $d_{x^2-y^2}$, and the $s_{xy}$ state. Their relative stability is determined mainly by the ratio of the strengths $V_1$ for nn and $V_2$ for nnn terms of the pair potential. A consistent interpretation of the experimental data can be achieved with a set of parameters for which the two states are almost degenerate. However, we find that the observed gap of electron doped materials can only be reproduced by an *attractive* on-site potential, which would indeed suggest that the pairing mechanism is different in this case.

We consider the following model Hamiltonian on a square lattice

$$H = \sum_{\mathbf{k},\sigma} \varepsilon(\mathbf{k}) c^\dagger_{\mathbf{k}\sigma} c_{\mathbf{k}\sigma} + H_{\text{int}} \quad , \tag{1}$$

$c^\dagger_{\mathbf{k}\sigma}$ creating a quasiparticle of momentum **k**, spin $\sigma$, where $H_{\text{int}}$ is chosen to have the following form

$$H_{\text{int}} = \sum_i \left[ V_0 \hat{n}_{i\uparrow}\hat{n}_{i\downarrow} + V_1 \sum_{\delta_n,\sigma,\sigma'} \hat{n}_{i\sigma}\hat{n}_{i+\delta_n,\sigma'} + V_2 \sum_{\delta_{nn},\sigma,\sigma'} \hat{n}_{i\sigma}\hat{n}_{i+\delta_{nn},\sigma'} \right] . \tag{2}$$

Here $\hat{n}$ stands for the number operator, $\delta_n(\delta_{nn})$ represents the nn (nnn) lattice vectors, and $\sigma$ is the spin index. The interaction constants $V_i$ are free parameters, however, generally we assume $V_0 \geq 0$, and $V_1, V_2 < 0$. This choice excludes isotropic s-wave pairing.

The quasiparticle dispersion relation $\varepsilon(\mathbf{k})$ is obtained by a tight binding fit to normal state ARPES data on Bi2212, adjusted so that $\varepsilon_F = 0$ (see Ref. [8] for details). Even though this fit was obtained from a particular material, we would like to stress that the qualitative features of the dispersion are *common* to all measured (by ARPES) high-$T_c$materials near optimal doping. The conclusions of this paper do not depend on the detailed form of $\varepsilon(\mathbf{k})$, as long as the dominant feature is incorporated, the presence of flat bands in a large region of the Brillouin zone, leading to a van Hove singularity (vHS) in the DOS (see Fig. 1). The singularity seems to be slightly ($\approx$ 30 meV for Bi2212, $\approx$ 20 meV for Y123) below $\varepsilon_F$ in the hole doped cuprates[10]. In the electron doped NCCO, flat bands are observed at $\approx 300$ meV below $\varepsilon_F$ [11]. A fit of $\varepsilon(\mathbf{k})$ to the latter spectra looks similar to the one obtained for the hole doped materials (taking into account a shift in the chemical potential).

**Table I**: The basis functions $\eta_i(\mathbf{k})$, their $C_{4v}$ group representations, and the corresponding coupling constants entering the pair potential $V(\mathbf{k}-\mathbf{k}') = \sum_{i=0}^{4} \widetilde{V}_i \eta_i(\mathbf{k})\eta_i(\mathbf{k}')$.

| $i$ | $\mathfrak{R}$ | $\eta_i(\mathbf{k})$ | | $\widetilde{V}_i$ |
|---|---|---|---|---|
| 0 | $A_1$ | 1 | $(s)$ | $V_0$ |
| 1 | $A_1$ | $\frac{1}{2}(\cos k_x + \cos k_y)$ | $(s^*)$ | $8V_1$ |
| 2 | $B_1$ | $\frac{1}{2}(\cos k_x - \cos k_y)$ | $(d_{x^2-y^2})$ | $8V_1$ |
| 3 | $A_1$ | $\cos k_x \cos k_y$ | $(s_{xy})$ | $8V_2$ |
| 4 | $B_2$ | $\sin k_x \sin k_y$ | $(d_{xy})$ | $8V_2$ |

After transforming $H_{\text{int}}$ to $\mathbf{k}$ space

$$H_{\text{int}} = \frac{1}{2\Lambda} \sum_{\substack{\mathbf{q},\mathbf{k},\mathbf{k}' \\ \sigma,\sigma'}} V(q)\; c^{\dagger}_{\mathbf{k}+\mathbf{q},\sigma} c^{\dagger}_{\mathbf{k}'-\mathbf{q},\sigma'} c_{\mathbf{k}',\sigma'} c_{\mathbf{k},\sigma} \;, \tag{3}$$

where $V(q) = V_0 + 4V_1(\cos q_x + \cos q_y) + 8V_2 \cos q_x \cos q_y$ and $\Lambda$ the volume, we apply standard BCS theory[12] to obtain the familiar gap equation

$$\Delta_{\mathbf{k}} = -\frac{1}{\Lambda} \sum_{\mathbf{k}'} V(\mathbf{k}-\mathbf{k}') \frac{\Delta_{\mathbf{k}'}}{2E_{\mathbf{k}}} \tanh\left(\frac{\beta E_{\mathbf{k}}}{2}\right). \tag{4}$$

Here $E_{\mathbf{k}} = \sqrt{\varepsilon_{\mathbf{k}}^2 + \Delta_{\mathbf{k}}^2}$ is the usual quasiparticle energy in the superconducting state, $\beta$ the inverse temperature. The potential $V(\mathbf{k}-\mathbf{k}')$ is separable, and restricting ourselves to singlet pairing, can be written as $V(\mathbf{k}-\mathbf{k}') = \sum_{i=0}^{4} \widetilde{V}_i \eta_i(\mathbf{k})\eta_i(\mathbf{k}')$. Each basis function $\eta_i(\mathbf{k})$ can be associated with one of the irreducible representations (in this case, all of them one-dimensional) of the $C_{4v}$ point group which characterizes the square lattice. They are listed in table II. The OP is expanded as $\Delta_{\mathbf{k}} = \sum_{i=0}^{4} \eta_i(\mathbf{k})$.

As usual, $T_c$ is determined by linearizing Eq. (4), which we can then solve separately within each irreducible representation. The representation with the largest $T_c$ is the one which will be chosen by the system, and as long as there is no additional phase transition, the OP at $T < T_c$ is given by the continous evolution of this solution as $T$ is lowered. Here, we shall not consider the consequences of a second transition.

The linearized gap equation becomes one-dimensional for the representations $B_1$ ($d_{x^2-y^2}$) and $B_2$ ($d_{xy}$), however for the $A_1$ case, we have mixing of the three basis functions $\eta_i(\mathbf{k}), i = 0, 1, 3$, and thus three coupled equations. In each irreducible representation $\mathfrak{R}$, the linearized gap equation can be written as

$$\Delta_i = -\frac{\widetilde{V}_i}{2\Lambda} \sum_{j \in \mathfrak{R}} \Delta_j \sum_{\mathbf{k}} \frac{\eta_i(\mathbf{k})\eta_j(\mathbf{k})}{\varepsilon_{\mathbf{k}}} \tanh\left(\frac{\beta_c \varepsilon_{\mathbf{k}}}{2}\right) \qquad (i \in \mathfrak{R}). \tag{5}$$

At this point it is instructive to define a new quantity for each basis function $\eta_i(\mathbf{k})$ as $P_i(\varepsilon) \equiv \frac{1}{\Lambda} \sum_{\mathbf{k}} \eta_i^2(\mathbf{k}) \delta(\varepsilon - \varepsilon_{\mathbf{k}})$, which we call its *pairing density of states* (PDOS). The role of the PDOS in the gap equation for an anisotropic pair state is analogous to the role of the usual single particle DOS, $N(\varepsilon)$, for an isotropic OP (for $\eta_0(\mathbf{k}) = 1$, $P_0(\varepsilon) = N(\varepsilon)$). The PDOS of a particular pair state is of crucial importance, since it is the value of $P_i(\varepsilon)$ around $\varepsilon_F$, which (apart from the pair potential strength) determines its stability.

In Fig. 1, we plot $P_i(\varepsilon)$ for the basis functions under consideration. This plot reveals an interesting and important point. The vHS in the DOS is reflected in the PDOS of only two of the four anisotropic basis functions, namely the $d_{x^2-y^2}$, and the $s_{xy}$ state. The $s^*$ state has almost vanishing weight, and the $d_{xy}$ state is also not very strong at $\varepsilon_F$. From this it is already clear that there are *two competing pair states* in the relevant doping range of the hole-doped cuprates. Their relative stability is determined mainly by the ratio $V_1/V_2$.

Now let us discuss in more detail the solutions of Eq. (5). Our strategy is as follows: first, we determine a set of realistic parameters for the coupling constants which can reproduce the measured $\mathbf{k}$ dependence of the gap in Bi2212, *i. e.*, an almost pure $s_{xy}$ gap. Second, we see, whether the same or slightly modified set of parameters gives results that are consistent with data from the other cuprates. We solve Eq. (5) numerically on a $d \times d$ $\mathbf{k}$-space grid, where the number $d$ at which convergence is obtained varies between 500 and 8000 depending on the size of the coupling constant. If the coupling constant is small, so is $T_c$, and one has to use more $\mathbf{k}$-points to accurately sample the important region around the FS, where the factor $\varepsilon_{\mathbf{k}}^{-1} \tanh(\beta\varepsilon_{\mathbf{k}}/2)$ is large. Note that the usual cutoff that prevents the logarithmic divergence in standard BCS theory is not necessary here due to the finite bandwidth.

The first criterion necessary to reproduce the Bi2212 gap is that an $A_1$ solution be stable versus the $d_{xy}$ state. Since the PDOS of the $s^*$ state is so small around $\varepsilon_F$, its component in the OP is negligible, unless we consider values $|V_1| \gg |V_2|$, for which the $d_{x^2-y^2}$ state is clearly the most favourable. Hence, in this comparison, we can safely set $V_1 = 0$. To gain some insight, we investigate the phase boundary between the two states in the $V_0, V_2$ plane. Figs. 2(a-c) show the phase boundaries for hole doping concentrations $\delta = 0.17 - 0.26$ (note that the vHS corresponds to $\delta = 0.34$). At $V_0 = 0$, we find that the $A_1$ state is favoured for $|V_2| \lesssim 200$meV. The stability of the $d_{xy}$ state at large coupling is explained as follows. The effective region being sampled in the $T_c$ equation grows as $T_c$ increases [this is easily understood from the $\beta_c$ dependence of the factor $\varepsilon_{\mathbf{k}}^{-1} \tanh(\beta_c \varepsilon_{\mathbf{k}}/2)$]. Hence, the contribution to the $d_{xy}$ pairing by the region around $\varepsilon = 0.15$, where its PDOS is large, grows substantially with increasing $|V_2|$. For $V_0 > 0$, the region of $A_1$ stability narrows to a $|V_2|$ stripe centered at about 100meV. As a function of increased hole doping, the phase boundary moves substantially towards larger $V_0$. This is a result of the increasing ratio of the $s_{xy}$ to the $d_{xy}$ PDOS $[P_3(\varepsilon_F)/P_4(\varepsilon_F)]$ as we approach the vHS (see Fig. 1). The region of realistic values for $T_c \lesssim 200K$ is obtained for $|V_2| \lesssim 100$meV regardless of $V_0$. This means that the $A_1$ solution is indeed stable versus the $d_{xy}$ state in the physically interesting parameter regime.

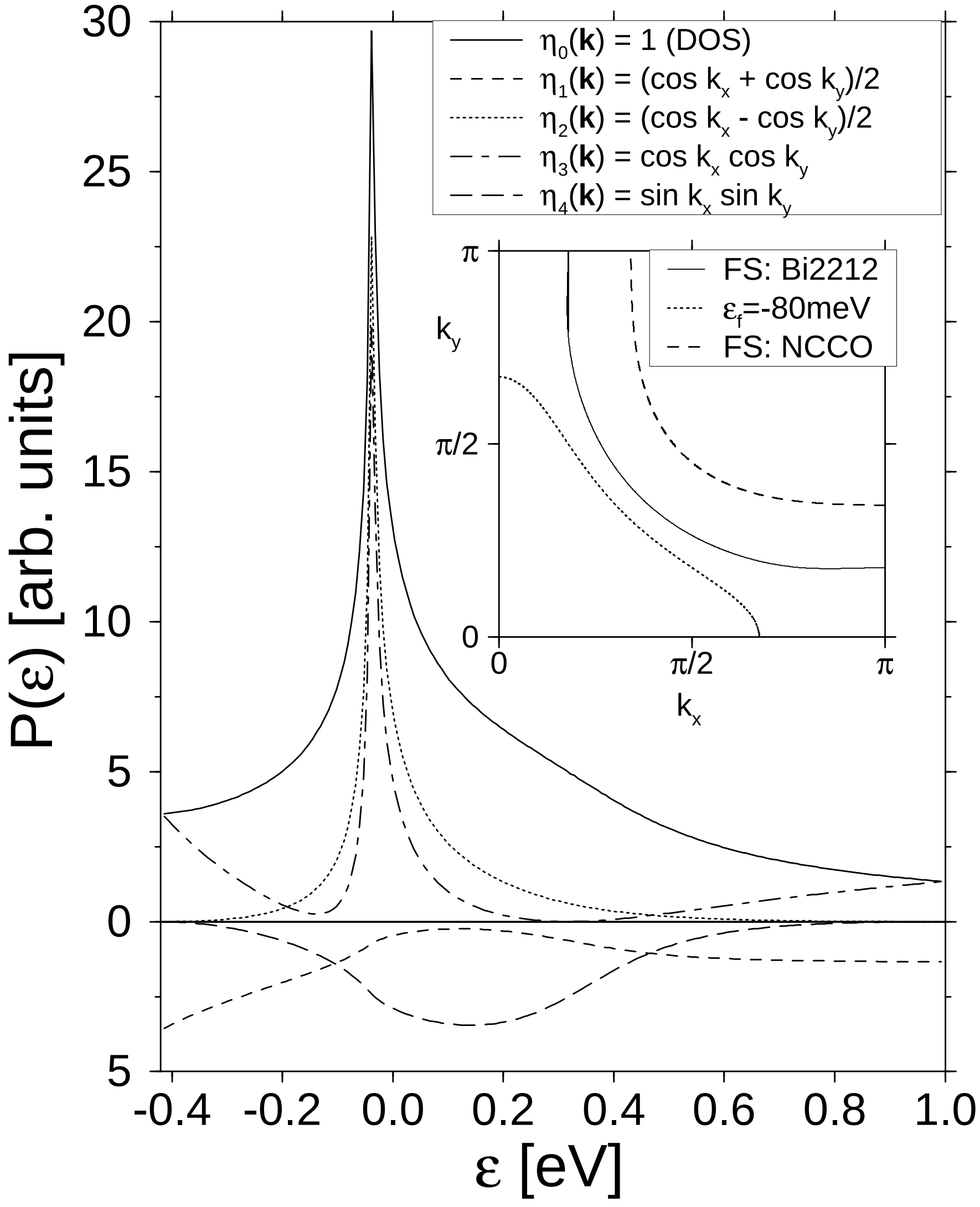

**Figure** 1 : The PDOS $P(\varepsilon)$ for the various OPs considered. In case of the isotropic gap $P(\varepsilon) = N(\varepsilon)$, the DOS. The inset shows the experimental Fermi surfaces for Bi2212, and NCCO as obtained from a tight binding fit to ARPES data. The wide region between the dotted and the continous line encloses the flat bands around the vHS ($\varepsilon_{\mathrm{vHS}} = -34$meV).

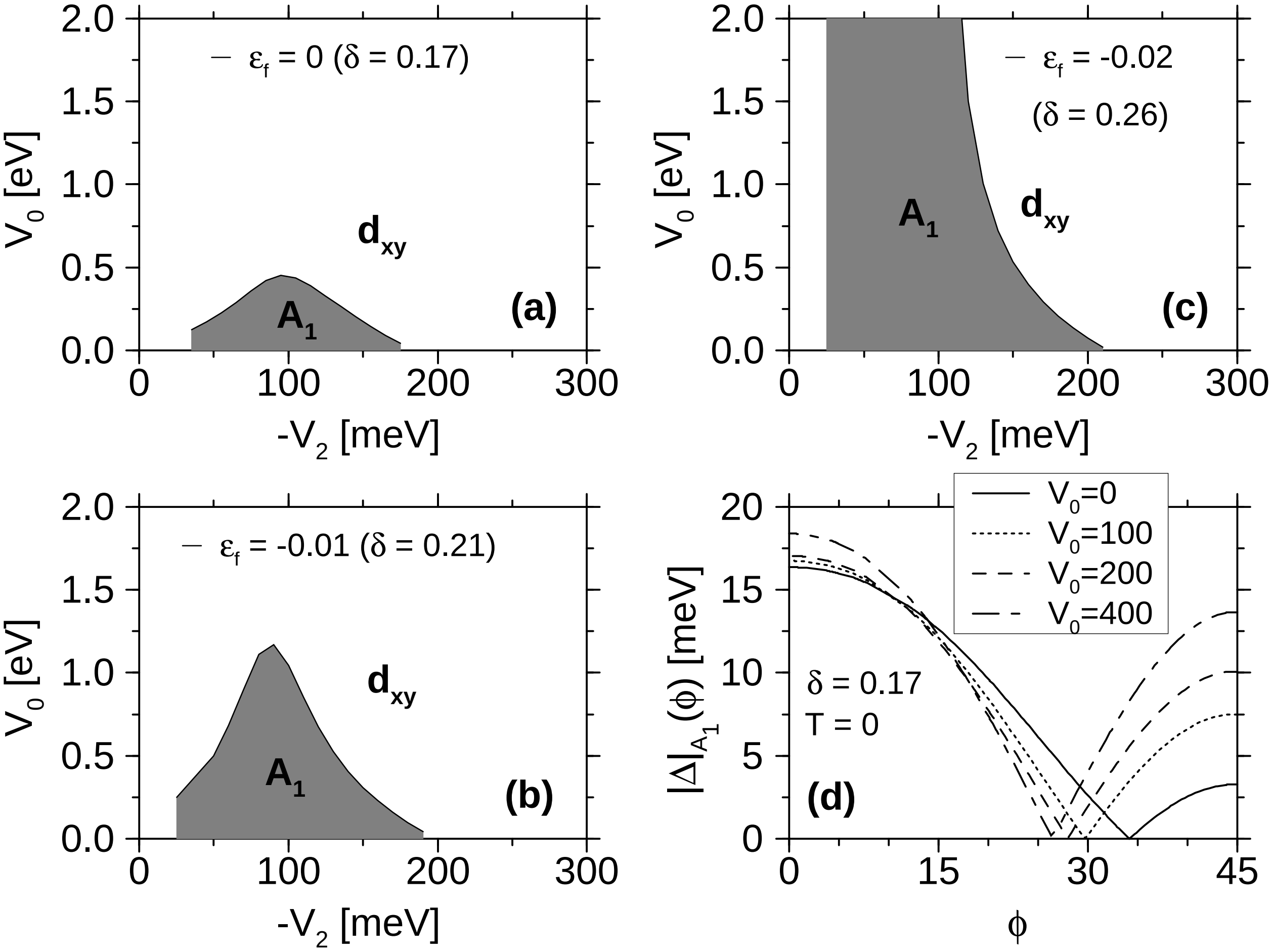

**Figure** 2 : (a-c) The phase boundary between the $d_{xy}$ and the most stable $A_1$ solution for three different doping levels ($V_1 = 0$). We did not determine the boundary, in the white region of small $|V_2|$, since $T_c < 1$K. (d) The angular dependence of the $T = 0$ superconducting gap in the $A_1$ representation for various strengths of the on-site repulsion $V_0$, calculated for the experimental FS of Bi2212. The coupling constants (in meV) were chosen as $V_1 = 0$, $(V_0, V_2) = [(0, -60), (100, -72), (200, -80), (400, -90)]$, which results in a value $T_c \approx 100$K for all four sets.

The second criterion that needs to be satisfied is the **k** dependence of the gap, in partic-ular, the position of its nodes, and its relative magnitude on either side of the nodes. Within the $A_1$ representation, this is determined by the ratio $V_0/V_2$. On general grounds, one can already anticipate the result: Since the penalty of a large Coulumb on-site repulsion can only be avoided if the two electrons making up a Cooper pair never visit the same site at the same time, one has to require that $\sum_i \langle c^\dagger_{i\uparrow} c^\dagger_{i\downarrow} \rangle = 0$. In BCS theory, this condition is equivalent to $\mathfrak{F}^\dagger(\mathbf{r}-\mathbf{r}' = 0, t = 0) \sim \sum_{\mathbf{k}} \Delta_{\mathbf{k}}/E_{\mathbf{k}} = 0$[12] ($\mathfrak{F}^\dagger$ the anomalous propagator), and requires a sign change of the gap, as well as roughly equal positive and negative portions of it in the region near the FS which contributes strongest to the sum. Hence for large $V_0$, we expect the nodes to converge to an angle near $\phi = \pi/8$ (22.5°). A quantitative analysis confirms this argument. In Fig. 3(d), we plot the $\phi$ dependence (measured from the $[\pi, \pi]$ point) of the superconducting gap at doping $\delta = 0.17$ (the experimental Bi2212 value) on the FS for various values of the on-site repulsion $V_0$. For each $V_0$, $V_2$ was chosen, so that $T_c \approx 100$K. For a value as low as $V_0 = 100$meV, we already find a shift of the node by 5°, and the size of the gap at 45° has more than doubled. Consequently, we have to conclude that the effective on-site repulsion between the quasi-particles has to be very weak to explain the observed gap in Bi2212 within an $A_1$ scenario. This is certainly a surprising result which requires a microscopic justification, since the bare copper on-site repulsion is generally believed to be large.

In the remainder of this paper, we concentrate on the competition between the $s_{xy}$ and the $d_{x^2-y^2}$ state. As a consequence of the previous considerations, we set $V_0 = 0$ in the calculations that follow. Fig. 3(a) shows the phase boundary between the two OPs in the relevant doping region. The boundary is essentially linear, and defined by a ratio of $V_2/V_1 \approx 1.4$. As a function of increasing hole doping there is a slight shift of the line towards smaller values of $|V_2|$. This means that in principle a transition from a $d_{x^2-y^2}$ to a $s_{xy}$ state is possible when the doping is increased provided that the two solutions are almost degenerate. Note, that this is again a straightforward consequence of the slightly different $\varepsilon$ dependence of the accompanying PDOS. To illustrate this point further, in Fig. 3(b) we plot the doping dependence of $T_c$ for the two states using two different sets of parameters. We note, that if the $s_{xy}$ state is stable, it is favoured only in the vicinity of the vHS. A transition to the $d_{x^2-y^2}$ state has to occur on either side of the optimal doping regime, provided $|V_1|$ is not extremely small. Figs. 3(c,d) show the coupling dependence of $T_c$ for the two competing states at two different doping levels $\delta = 0.17, 0.26$. At large coupling the dependence becomes approximately linear in $V_1, V_2$ resp.[13], a consequence of the proximity to the vHS. For $\beta_c \gtrsim 20 - 30$meV the integrated PDOS over the region close to $\varepsilon_F$, where $\varepsilon_{\mathbf{k}}^{-1} \tanh(\beta_c \varepsilon_{\mathbf{k}}/2) \approx \beta_c/2$, essentially saturates as a function of increasing $\beta_c$, since it is exhausted by the vHS (see Fig. 1). Hence the $\beta_c$ dependence of the sum in Eq. (5) becomes approximately linear.

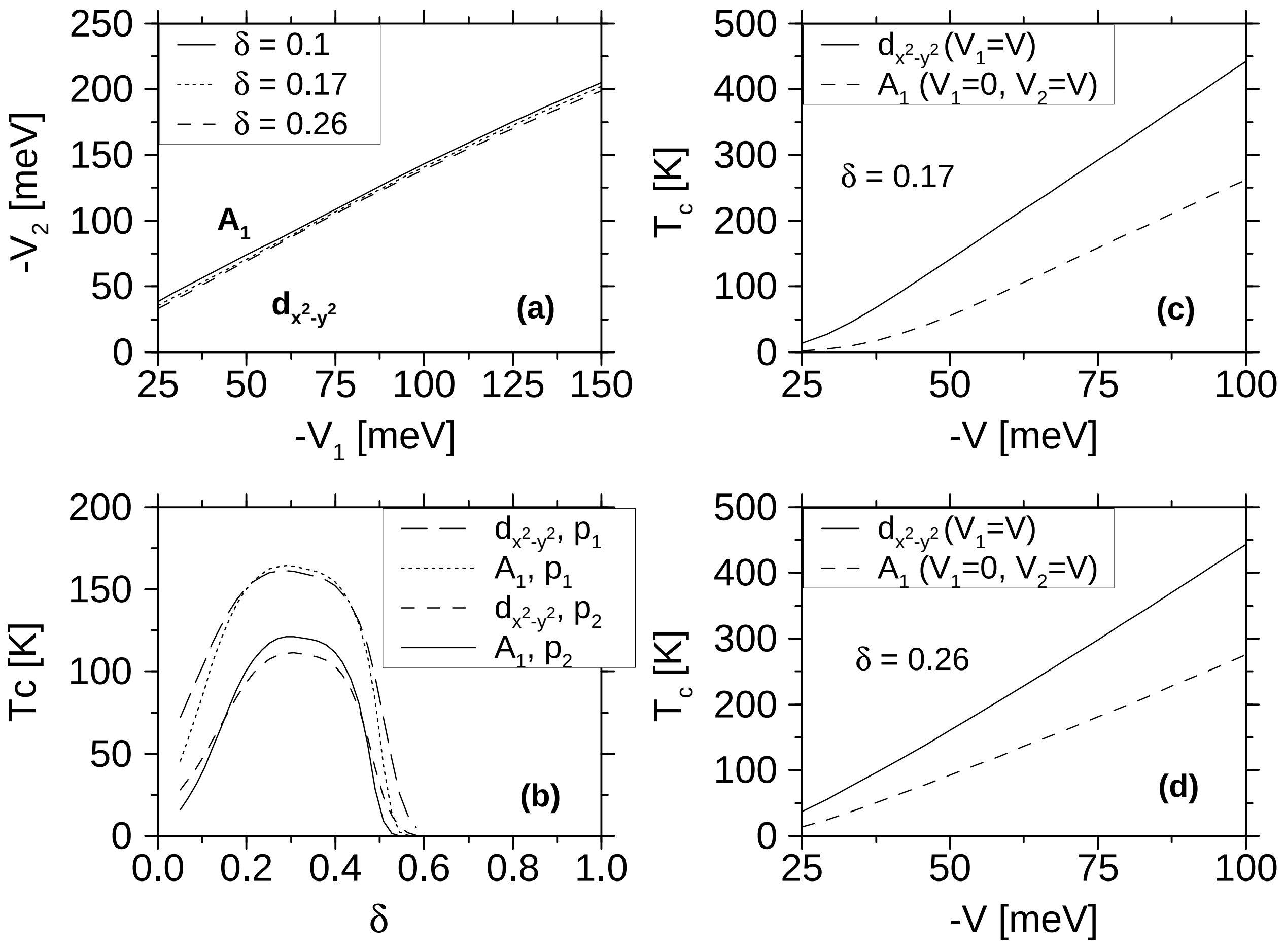

**Figure** 3 : (a) The phase boundary between the $d_{x^2-y^2}$ and the most stable $A_1$ solution for three different doping levels ($V_0 = 0$). (b) The doping dependence of $T_c$ for the $d_{x^2-y^2}$ and the $A_1$ state calculated using two different sets of parameters, $p_1, p_2$ given by $(V_1, V_2) = [(-50, -70), (-40, -58)]$ (in meV). (c,d) The coupling dependence of $T_c$ for the two states at two different doping levels $\delta = (0.17, 0.26)$.

Until now, we were mainly concerned about the determination of a set of parameters which can reproduce the observed gap in Bi2212. In the final part of the paper, we want to see whether a similar set of parameters can consistently explain the experimental data relevant to the OP symmetry in Y123. As already noted by other authors [14], the lower orthorhombic symmetry of Y123 has the effect that the five basis functions we considered now belong to only two different irreducible representations of the relevant point group $C_{2v}$. In particular, the $d_{x^2-y^2}$ state now falls into the identity representation $A_1$, and a mixing with the s-wave like basis functions at $T_c$ is now possible. Note, that even though the true symmetry of Bi2212 is also orthorhombic due to the superlattice modulation, since the latter occurs along the (1,1) direction, the $d_{x^2-y^2}$ state does not belong to the $A_1$ representation in this case, and hence no mixing is possible[8]. To simulate the orthorhombic distortion, we artificially introduce a difference in the $k_x$ versus $k_y$ quasiparticle dispersion characterized by a parameter $\alpha$ defined by setting $t_{i,x} = (1-\alpha)t_i, t_{i,y} = (1+\alpha)t_i$. Here $t_i$ is the real space hopping matrix element for a lattice vector $\delta_i$ in the tight-binding formulation. The qualitative effect of this symmetry breaking does not depend much on the value of $\alpha$, so we fix it at $\alpha = 0.05$. In our model, the orthorhombic distortion splits the vHS into two peaks, roughly separated by 50meV. The peak at higher energy corresponds to doping $\delta = 0.24$. Note, that this splitting is of no relevance to the arguments presented below.

Since we already learned that the two competing states in the hole-doped region are the $d_{x^2-y^2}$ and the $s_{xy}$ state, in the following analysis, we restrict ourselves to the mixing of those two states, *i. e.*, we write the gap function as $\Delta_{\mathbf{k}}(T) = \Delta_1(T)\eta_2(\mathbf{k}) + \Delta_2(T)\eta_3(\mathbf{k})$, and now we solve the full non-linear gap equation to determine the coefficents $\Delta_1(T), \Delta_2(T)$. In the parameter regime previously determined (where $d_{x^2-y^2}$ and $s_{xy}$ are almost degenerate) , we find considerable mixing between the two states. In Figs. 4(a,b), we show the angular dependence of the $T = 0$ OP at two different doping levels $\delta = 0.24, 0.19$ for various values of the coupling constants $V_1, V_2$ going from the case of a dominating $s_{xy}$ to a dominating $d_{x^2-y^2}$ component. The crucial result here is that at $T = 0$, the two solutions repel each other, *i. e.*, there is no value of the coupling constants where $\Delta_1 \approx \Delta_2$. The maximum admixture of the minor component is roughly 25%. This is clearly demonstrated in Fig. 4(d), where the temperature dependence of the two components is plotted for two sets of parameters. For $T$ close to $T_c$, the two components develop roughly equally with decreasing $T$ until at some $T$, one of the components starts to be boosted, and the other one becomes suppressed. This cross-over temperature can be related to the crossing of the free energy of the two solutions in tetragonal symmetry, which can occur when they have a similar $T_c$. If the solutions are in the same representation, however, a crossing cannot occur; the solutions come very close to each other, and at the cross-over temperature, the stable solution changes its character rapidly, but continously.

This behaviour can lead to an interesting effect, which is demonstrated in Fig. 4(c). It is actually possible that as a function of $T$, the solution changes so dramatically, that close to $T_c$, the sign of the gap is the same at 0° and 90°, whereas as $T$ is lowered, the $d_{x^2-y^2}$ character starts to dominate, and the gap changes sign under a 90° rotation. Experimental evidence for such behaviour has already been reported on samples of Y123 [15].

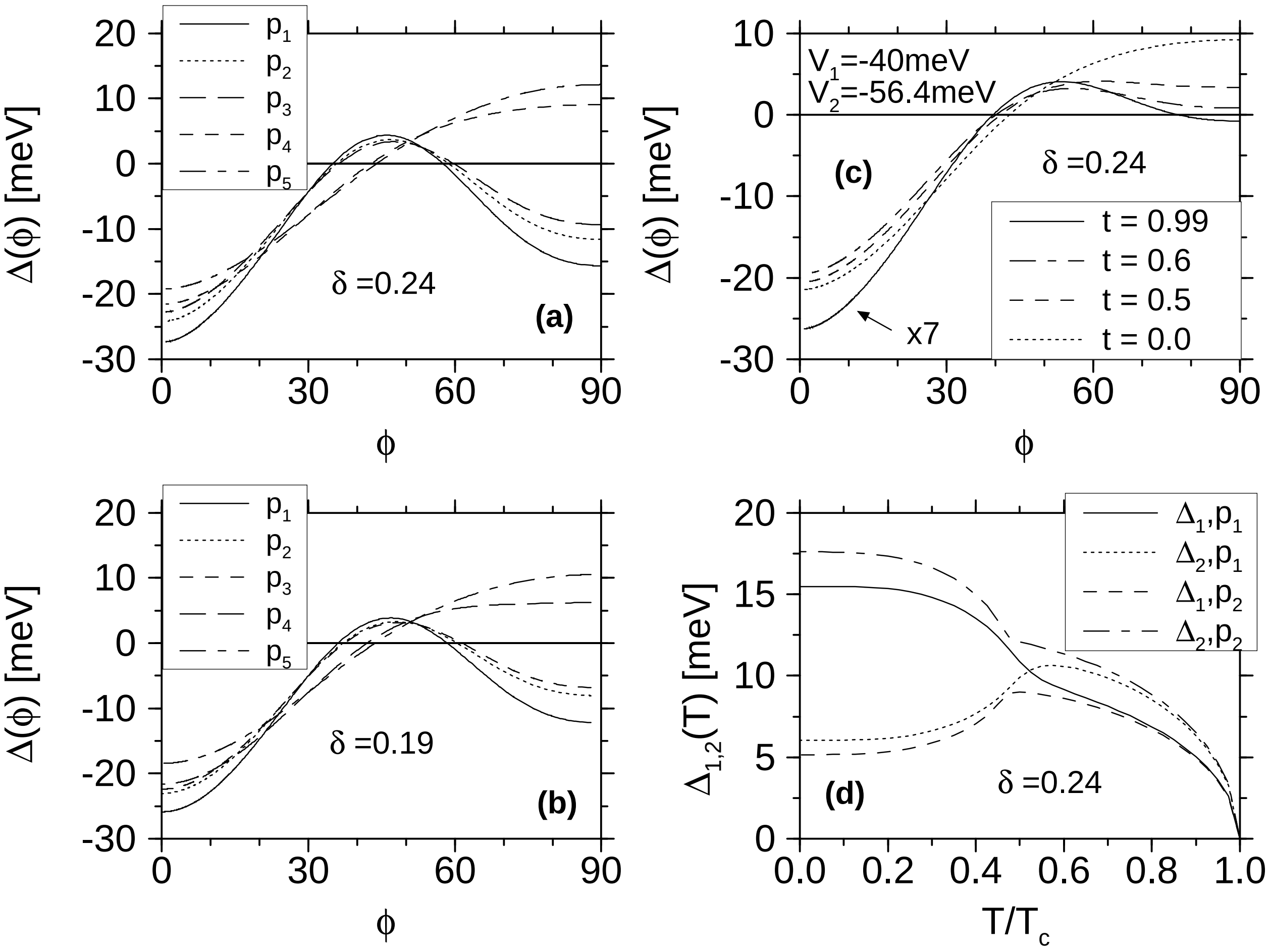

**Figure** 4 : Effect of the orthorhombic distortion: (a,b) The angular dependence of the $T = 0$ mixed $d_{x^2-y^2} + s_{xy}$ OP in orthorhombic symmetry at two doping levels for different sets of parameters, $\delta = 0.24$, $p_1 \cdots p_5$ defined by $V_1 = -40$, $V_2 = [-65, -60, -57.6, -57.2, -40]$, and $\delta = 0.19$, $p_1 \cdots p_5$ defined by $V_1 = -40$, $V_2 = [-65, -60, -58.8, -58.5, -40]$. (c) The angular dependence of the mixed $d_{x^2-y^2} + s_{xy}$ OP at different temperatures ($t = T/T_c$). (d) The temperature dependence of the $d_{x^2-y^2}$ component $\Delta_1$, and the $s_{xy}$ component $\Delta_2$ of the mixed $d_{x^2-y^2} + s_{xy}$ OP for two different sets of parameters, $p_1, p_2$ given by $V_1 = -40$, $V_2 = [-57.5, -57]$.

In any case, for the set of parameters determined to fit the gap of Bi2212, we find considerable mixing of the $d_{x^2-y^2}$ and the $s_{xy}$ state in Y123. As shown, it is not difficult to obtain a gap function with predominantly $d_{x^2-y^2}$ character so that the OP changes sign under a 90° rotation, while its average over the FS is still considerably large due to the $s_{xy}$ component. This could resolve the controversy which arose from results of different phase sensitive measurements on Y123 mentioned at the beginning.

In conclusion, we have shown that a phenomenological model based on standard BCS theory can consistently explain the experimental data concerning the pair state of the hole doped cuprates. Three ingredients are required to make this interpretation possible: (i) The existence of flat bands (ii) a very small effective $V_0$ (iii) a slightly dominating effective nnn attraction of the order of 60-80meV with a ratio $V_2/V_1 \approx 1.3 - 1.5$. Clearly a derivation of the effective model presented here from a Hamiltonian with realistic, physically motivated parameters and interactions is an important next step in the understanding of the true pairing mechanism. Finally, we just briefly mention that the electron doped NCCO does *not* fit into this scenario. Given the experimental quasiparticle dispersion we found that only isotropic s-wave pairing, *i. e.*, a negative $V_0 \approx -200$meV, can reproduce the experimentally observed nodeless superconductivity. For the anisotropic states considered, the $d_{x^2-y^2}$ state is by far the most stable in this case.

We acknowledge financial support by the National Science Foundation (NSF-DMR-91-20000) through the Science and Technology Center for Superconductivity (R.F.), and of the U.S. Department of Energy, Office of Basic Energy Sciences, under Contract No. W-31-109-ENG-38 (M.R.N.).